\documentclass{sig-alternate-05-2015}
\CopyrightYear{2017} 
\setcopyright{rightsretained}
\conferenceinfo{To appear in Proceedings of CHIIR 2017}{}
\isbn{978-1-4503-4677-1/17/03}\acmPrice{\$15.00}
\doi{http://dx.doi.org/10.1145/3020165.3022129}
\usepackage[font=small,skip=3pt]{caption}
\begin{document}



%

\title{Personalised Query Suggestion for Intranet Search with Temporal User Profiling}
%
%
%
%
%

%
\author{
%
%
   \alignauthor
	Thanh Vu$^1$, Alistair Willis$^1$, Udo Kruschwitz$^2$, and Dawei Song$^{1,3}$\\
       \affaddr{$^1$ The Open University, Milton Keynes, United Kingdom\\ $^2$ University of Essex, Essex, United Kingdom\\$^3$ Tianjin University, Tianjin, P.R.China}
       \email{\footnotesize \{thanh.vu, alistair.willis, dawei.song\}@open.ac.uk, udo@essex.ac.uk}      
}

\maketitle
\begin{abstract}
Recent research has shown the usefulness of using collective user interaction data (e.g., query logs) to recommend query modification suggestions for Intranet search. However, most of the query suggestion approaches for Intranet search follow an ``one size fits all'' strategy, whereby different users who submit an identical query would get the same query suggestion list. This is problematic, as even with the same query, different users may have different topics of interest, which may change over time in response to the user's interaction with the system.

We address the problem by proposing a personalised query suggestion framework for Intranet search. For each search session, we construct two temporal user profiles: a \emph{click user profile} using the user's clicked documents and a \emph{query user profile} using the user's submitted queries. We then use the two profiles to re-rank the non-personalised query suggestion list returned by a state-of-the-art query suggestion method for Intranet search. Experimental results on a large-scale query logs collection show that our personalised framework significantly improves the quality of suggested queries.
\end{abstract}

{\small\textbf{Categories and Subject Descriptors:} H.3.3 [Information Systems Applications]: Information Search and Retrieval}


\section{Introduction}
Query suggestion is an important feature in web search engines (e.g., Bing, Google) as well as in domain-specific search engines (e.g., Intranet search) \cite{Adeyanju12}. Query suggestions help users quickly refine the input query to better meet the user's information need by recommending possible terms to modify the original input query.

In this paper, we focus on query suggestion for Intranet search, which is different from web search \cite{Hawking10}. Specifically, Intranet search (e.g., university intranets) is \emph{domain-specific} and built to satisfy the user's information need related to a specific domain (e.g., the university's document corpus). Moreover, the Intranet may not be fully indexed and accessible by web search engines. For example, web search engines cannot access those Intranet documents which require authorised logins. The searcher, therefore, may need to use an Intranet search engine to locate relevant documents.

Using collective user interaction data (e.g., query logs) for query suggestions has been shown useful for Intranet search \cite{Adeyanju12, Albakour2011}. Existing Intranet search approaches appear to follow a ``one size fits all'' strategy. That is different users who submit the same query receive the same query suggestion list. However, different users may have different topics of interest. Consequently, the users who have submitted the same query may have different search intentions. For example, a sociology student submitting the query ``lecture notes'' is likely to be more interested in sociology classes than maths classes. Moreover, users' interests and search intentions may be dynamically evolving depending on their interactions with the system (e.g., clicks on documents), and when the interactions are made, during a search session \cite{BennettM12}. 

To address these issues, we propose a unified framework to personalise query suggestions for Intranet search. Specifically, we use the interaction data of each user during a search session to build \emph{user profiles}, which represent the user's topics of interests and may change over time in response to the user's interaction with the system. 

It is worth noting that search personalisation (e.g., search result re-ranking, query suggestion, query auto-completion, etc.) has been studied extensively in the context of web search engines \cite{BennettM12, ShokouhiL2013, Shokouhi2013, Vu14, VuW15, Vu15, Vu17}. However, little attention has been paid to the same task for Intranet search. Moreover, personalisation methods on web search engines typically construct users' profiles using their click information \cite{BennettM12, Vu14, VuW15, Vu15}, but less account has been taken of how users modify their queries for building the profiles. 

In our proposed framework, we construct two temporal topic-based user profiles for each search session. The first is a \emph{click user profile} based on the clicked documents. The second is a \emph{query user profile} based on the user's query modification history within the search session. We then use the two profiles within a learning-to-rank framework to re-rank suggested queries generated by a non-personalised method for query suggestion on Intranet search \cite{Adeyanju12}. Experimental results show that our approach helps to significantly improve the query suggestion performance.

\section{Query Suggestion Framework}
\subsection{Building Temporal User Profiles}
\label{sec21}
Each search session contains two types of event (i.e., queries and clicks). Given a search session, we propose to build two temporal profiles for the specific user. These are a click user profile (denoted as $profile(C)$), built using the user's clicked documents, and a query user profile (denoted as $profile(Q)$), built using the user's submitted queries within the session. Click profile have been extensively used in other search personalisation methods \cite{BennettM12}; we expect that the query profile will enrich the representation of the user's search interests.

Since a user's interests and search intentions may change over time, the more recently clicked documents and submitted queries could better represent the user's current interests. In this paper, we propose to use a \emph{decay function} to capture this characteristic as in \cite{BennettM12,Vu15}.

\subsubsection{Extracting Topics from Clicked Documents}
We consider that click information (e.g., clicked documents) is a good indicator of those documents' relevance to the user's interests \cite{Joachims2005}. To build the user profile, we use the topics discussed in the documents. We first extract clicked documents from the Intranet search's query logs. After that, we employ Latent Dirichlet Allocation (LDA) \cite{BleiL03} to automatically extract latent topics (denoted as $Z$) from the clicked documents (denoted as $D$). After training an LDA model using the clicked documents, we apply the model to extract topics for the remaining documents in the collection. Finally, each document is described as a multinomial distribution over the topics (denoted as $P(Z|D)$), in which each topic is represented as a multinomial distribution over the entire vocabulary.

\subsubsection{Building a Click User Profile}
We represent the temporal click user profile as a multinomial distribution over the topics as in \cite{BennettM12}. Specifically, the user set is denoted as $U$. Let $u$ be an instance of $U$. Let $D_c = \{d_{c_1}, d_{c_2}, ..., d_{c_n}\}$ be the set of clicked documents of the user $u$ in the current search session, we define the click user profile of the user $u$ (given the clicked document set $D_c$) as a distribution over topics $Z$ (denoted as $P_C(Z|U)$). The probability $p_C(z|u)$ indicates how much the user $u$ is interested in topic $z \in Z$. $p_C(z|u)$ is defined as a mixture of probabilities of $z$ given $d_{c_i} \in D_c$ as follows
{\small
\begin{equation}
\label{eq1}
\begin{aligned}
p_C(z|u) = \sum\nolimits_{d_{c_i} \in D_c} \lambda_ip(z|d_{c_i})
\end{aligned}
\end{equation}
}
$\lambda_i = \frac{1}{N}\alpha^{t_{d_{c_i}} - 1}$ is the exponential decay function of $t_{d_{c_i}}$, which is the order of the document $d_{c_i}$ clicked by the user $u$ in the search session. $t_{d_{c_i}} = 1$ indicates that $d_{c_i}$ is the most recently clicked document; $N$ is the normalisation factor. $\alpha$ is the decay parameter ($0 \leq \alpha \leq 1$). 
\subsubsection{Building a Query User Profile}
Let $Q = \{q_1, q_2, …, q_m\}$ be the submitted query set of $u$ in the search session. Because the number of Intranet documents is smaller and can be assumed to change less frequently than web search engines', we make the simplifying assumption of describing each query by the set of documents that contain all the query words, denoted as $D_{q_i} = \{d_{i_1}, d_{i_2}, ..., d_{i_k}\}$. Then, each search query $q_i$ (given the document set $D_{q_i}$) is modelled as a distribution over topics $Z$ (denoted as $P(Z|q_i)$). The probability of a topic $z \in Z$ given $q_i \in Q$ (i.e., $p(z|q_i)$) is defined as a mixture of probabilities of $z$ given a document $d_{i_j} \in D_{q_i}$ as follows
{\small
\begin{equation}
\label{eq2}
\begin{aligned}
p(z|q_i) = \sum\nolimits_{d_{i_j} \in D_{q_i}} \frac{1}{|D_{q_i}|}p(z|d_{i_j})
\end{aligned}
\end{equation}
}
$|D_{q_i}|$ is the size of the document set $D_{q_i}$. 

We then model the query user profile of the user $u$ (given the query set $Q$) as a distribution over topics $Z$ (denoted as $P_Q(Z|u)$). The probability of a topic $z$ given $u$ (i.e., $p_Q(z|u)$) is defined as a mixture of probabilities of $z$ given query $q_i \in Q$ as follows
{\small
\begin{equation}
\label{eq3}
\begin{aligned}
p_{Q}(z|u) = \sum\nolimits_{q_i \in Q}\lambda_ip(z|q_i)
\end{aligned}
\end{equation}
}

$p(z|q_i)$ is defined in Equation \ref{eq2}. Similar to the click user profile, $\lambda_i = \frac{1}{M}\alpha^{t_{q_i} - 1}$  is the exponential decay function of $t_{q_i}$, which is the order of the query $q_i$ submitted by the user $u$ in the search session.
 $t_{q_i} = 1$ indicates that $q_i$ is the most recent query; $M$ is the normalisation factor. $\alpha$ is the decay parameter ($0 \leq \alpha \leq 1$).
\subsection{Re-ranking Suggested Queries}
\label{sec23}
We use the two user profiles in a learning-to-rank mechanism to re-rank the query suggestion list returned by a non-personalised query suggestion method proposed by Adeyanju \emph{et al.} \cite{Adeyanju12}, denoted as \emph{Adeyanju's}. Specifically, \emph{Adeyanju's} first constructs a domain knowledge structure in the form of a concept subsumption hierarchy using both the Intranet document collection and collective users' query logs. Next, the suggestion list is generated using the top $n$ terms most relevant to the query in the hierarchy. 

For each input query, our re-ranking method is detailed as follows

\textbf{(1)} We generate the top $n$ ranked suggested queries using \emph{Adeyanju's} method. We denote a suggested query as $q_s$.

\textbf{(2)} We then compute similarity scores between $q_s$ and $profile(C)$, and between $q_s$ and $profile(Q)$. Both the suggested query $q_s$ and a user profile (denoted as $pf$ which is either $profile(C)$ or $profile(Q)$) are modelled as distributions over topics $Z$ (Section \ref{sec21}). To measure the similarity between $q_s$ and the user profile $pf$, we use Jensen-Shannon divergence ($D_{JS}\lfloor.||.\rfloor$), which is a popular method of measuring the divergence (similarity) between two distributions, to measure the similarity between $q_s$ and $pf$
{\small
\begin{equation}
\label{eq:9}
Sim(q_s|pf) = -D_{JS}\lfloor Q||P\rfloor = -\big(\frac{1}{2}D_{KL}\lfloor Q||M\rfloor + \frac{1}{2}D_{KL}\lfloor P||M\rfloor\big)
\end{equation}
}
Here, $Q$ and $P$ are distributions over topics of $q_s$ and $pf$, respectively. $D_{KL}\lfloor .||.\rfloor$ is the Kullback-Leibler divergence and $M = \frac{1}{2}(Q+P)$. We consider the scores as the \emph{personalised features}. We also extract other \emph{non-personalised features} of the input query $q$ and the suggested query $q_s$. Table \ref{tb1} shows the features extracted for re-ranking the suggestion list.

\vspace{3pt}
\begin{table}[ht]
\centering
\caption{\small{The personalised query suggestion features}}
\resizebox{8.5cm}{!}{
\begin{tabular}{ll}\hline
\textbf{Feature} & \textbf{Description}\\\hline
\multicolumn{2}{l}{\textbf{Personalised Features}}\\\hline
ClickPersonalisedScore
 & The similarity score between the suggested query and the user click profile\\
 QueryPersonalisedScore
 & The similarity score between the suggested query and the user query profile\\\hline
\multicolumn{2}{l}{\textbf{Non-personalised Features}}\\\hline
QueryRank & Rank of the suggested query on the original list\\
QuerySim & The cosine similarity score between the current query and the previous query\\
QueryNo & Total number of queries that have been submitted to the Search Engine\\
SuggestedQueryCosine & The cosine similarity score between the current query and the suggested query\\
SuggestedQueryJaccard & The Jaccard distance score between the current query and the suggested query\\
SuggestedQueryEdit & The edit distance between the current query and the suggested query\\
SuggestedQueryLevenshtein & The Levenshtein distance between the current query and the suggested query\\
SuggestedQueryPreUsed & Whether the suggested query was used by the user in the same search session?\\
\hline
\end{tabular}
}
\label{tb1}
\end{table}

\textbf{(3)} After extracting the query features, to re-rank the top $n$ suggested queries, we employ LambdaMART \cite{Burges07} to train ranking models. Among many learning-to-rank algorithms, LambdaMART is regarded as one of the best-performing algorithms and has been chosen as the base learning algorithm in various recent approaches to search personalisation \cite{BennettM12}. 
\newpage
\section{Experimental Methodology}
\label{sec3}
\subsection{Evaluation Methodology and Dataset}
\textbf{Evaluation methodology} For evaluation, we use \emph{AutoEval}, an automated evaluation framework, which measures the performance of query suggestions automatically based on the actual query logs of an Intranet search \cite{Albakour2011}. For each query \emph{suggestion list}, we assign a positive label for a suggestion if it is an actual \emph{refinement}, which is the next submitted query in the search session, and there is at least one user click on retrieved results after the refinement. In other words, we interpret the user click after a reformulation as the criterion of a relevant suggestion. The remainder of the suggestion list is assigned negative (irrelevant) labels. We use the rank positions of the positively labelled queries as an approximation of the ground truth to evaluate the performance of query suggestions before and after re-ranking. 

We also follow the experimental methodology in \cite{Adeyanju12}, that is, the model is evaluated continuously at periodic intervals. Specifically, we use the logs in week $i$ for training the re-ranking model and the following week $i+1$ for testing the trained model; where, in our experiments, $1 \leq i \leq w$, the number of weeks in the test period.

\textbf{Dataset} The dataset used in our experiments contains large-scale query logs collected from the search engine installed at the Web site of the University of Essex during the two years covering 1 January 2012 - 31 December 2013. Each log sample contains a session identifier, the event type (i.e., a query or a click), an auto-increment id, the event content (i.e., query text, click URL), and the event time-stamp.

We apply a simple pre-processing step to remove single event search sessions because, with those sessions, it is not possible to determine whether the user found the required information. We also remove those queries whose positive label set is empty from the dataset. We then analyse the remainder of the query logs. Table \ref{tb2} shows basic statistics.

\begin{table}[ht]
\centering
\caption{Basic statistics of the evaluation search logs}
\label{tb2}
\resizebox{!}{1.2cm}{
\begin{tabular}{c c c c} \hline
Item & 2012 & 2013 & Total\\\hline
\#search sessions & 397,461 & 338,391 & 735,804\\\hline
\#events & 1,263,179 & 1,083,992 & 2,347,171\\ \hline
\textbf{\#events/session}& \textbf{3.22} & \textbf{3.25} & \textbf{3.23}\\\hline
\#queries & 757,645 & 659,284 & 1,416,929\\\hline
\textbf{\#query/session} & \textbf{1.91} & \textbf{1.95} & \textbf{1.93}\\\hline
\#clicked url & 505,534 & 424,708 & 930,242\\\hline
\textbf{\#clicks/session} & \textbf{1.27} & \textbf{1.26} & \textbf{1.26}\\\hline
\end{tabular}
}
\end{table}
\vspace{-6pt}
\subsection{Experimental Settings}
\label{sec32}
\textbf{Our personalisation method and baselines} We name our proposed re-ranking model as \emph{Ours}. We choose two baselines to compare our work against. The \emph{first baseline} is \emph{Adeyanju's} method \cite{Adeyanju12}, which we reimplemented to generate the original suggestion list for \emph{re-ranking}. 

We use the session-based approach proposed by Bennett \emph{et al.} \cite{BennettM12} as the \emph{second baseline}. Specifically, in the baseline, we use only the click user profile (i.e., $profile(C)$) together with the non-personalised features detailed in Table \ref{tb1} to re-rank the suggestion list. We name the baseline as \emph{Click}. 

It is worth noting that \emph{Adeyanju's} is a non-personalised
approach and achieved a good performance of query suggestion
on Intranet search \cite{Adeyanju12}. \emph{Click} is a personalised approach
and achieved good performances in web search personalisation \cite{BennettM12}. Moreover, instead of the session-based approach by Bennett \emph{et al.} \cite{BennettM12} as our personalised baseline we could alternatively have used Shokouhi \emph{et al.} \cite{Shokouhi2013} or Shokouhi \cite{ShokouhiL2013}.

\textbf{LDA \& LambdaMART} We train an LDA model on the clicked documents extracted from the query logs, as detailed in Section \ref{sec21}. The number of topics (i.e., 300 in our experiments) is decided by using a held-out validation set which consists of 10\% of all clicked documents. The selected number of topics is the one that gives the lowest perplexity value. The decay parameter $\alpha$ for the two user profiles is set to $0.95$ as in \cite{BennettM12}. The ranking function is learned using LambdaMART \cite{Burges07}. We used the default setting for LambdaMART's prior parameters\footnote{Number of leaves = 10, minimum documents per leaf = 200, number of trees = 100 and learning rate = 0.15}.

\textbf{Evaluation metrics} The evaluation is based on the comparison between our personalised approach and the baselines. We use four evaluation metrics: Mean Average Precision (MAP), Precision (P@k), Mean Reciprocal Rank (MMR), and Normalized Discounted Cumulative Gain (nDCG@k).

\section{Results}
\label{sec4}
\subsection{Overall Performance}
Table \ref{tb3} shows promising results when user profiles are used to personalise the query suggestion list. We can see that, even using only the click user profile, the \emph{Click} method has led to an improvement of 7.22\% on MAP over \emph{Adeyanju's} method. The combination of query and click profiles (i.e., \emph{Ours}) achieves the highest improvement of 10.97\% over \emph{Ade\-yanju's} in terms of MAP score. The improvements indicate that personalisation helps improve the query suggestion performance. The improvements over \emph{Adeyanju's} are all significant (\emph{paired t-test}, p < 0.01).

\vspace{3pt}
\begin{table}[ht]
\centering
\caption{Overall performance of the methods. \%rel denotes the relative improvement over \emph{Adeyanju's}.}
\label{tb3}
\resizebox{8cm}{!}{
\begin{tabular}{c c c c c c c} \hline
Model & MAP & P@1 & P@5 & MRR@10 & nDCG@5 & nDCG@10\\\hline
Adeyanju's & 0.5440 & 0.4113 & 0.1823 & 0.5447 & 0.5714 & 0.6000\\\hline
Click & 0.5833 & 0.4271 & 0.1981 & 0.5839 & 0.6193 & 0.6583\\
\%rel &{\small \emph{+7.22\%}}&{\small \emph{+3.84\%}}&{\small \emph{+8.67\%}}&{\small \emph{+7.22\%}}&{\small \emph{+8.38\%}}&{\small \emph{+9.73\%}}\\\hline
\textbf{Ours} & \textbf{0.6037} & \textbf{0.4526} & \textbf{0.2026} & \textbf{0.6043} & \textbf{0.6413} & \textbf{0.678}\\
\%rel &{\small \emph{+10.97\%}}&{\small \emph{+10.04\%}}&{\small \emph{+11.14\%}}&{\small \emph{+10.94\%}}&{\small \emph{+12.23\%}}&{\small \emph{+13.02\%}}\\\hline
\end{tabular}
}
\end{table}

In the comparison between personalisation methods (i.e., \emph{Ours} and \emph{Click}), Table \ref{tb3} shows that using both the click and query user profiles (i.e., \emph{Ours}) significantly improves the suggestion quality over the \emph{Click} baseline ($p < 0.01$). Interestingly, our method produces a significantly better quality of the first query in the suggestion list with the improvement of 5.97\% on P@1 over \emph{Click}. The improvements of \emph{Ours} over \emph{Click} also indicate that the query user profile is important in the query suggestion task, especially for the quality of the first suggested query.
\subsection{Performance on Different Query Positions}
With more submitted queries and clicked documents, we are able to build richer user profiles. In this experiment, we aim to study whether the position of a query in a search session has any effect on the performance of personalised query suggestion. For each search session, we label queries by their positions during the session. Because there are few sessions containing more than three queries (i.e., 7.65\% of sessions in the query logs), we label the first three queries from one to three according to the order of submission in the search session; the remaining queries are labelled as $\geq 4$.

\begin{figure}
\centering
\includegraphics[width=6.0cm]{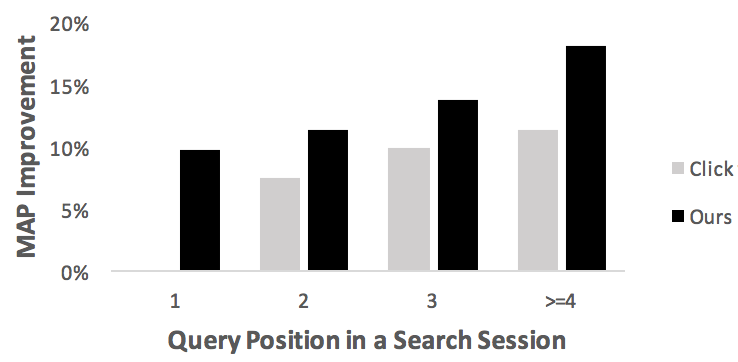}
\caption{Relative MAP improvements over \emph{Adeyanju's}  with different query positions. 
}
\label{img1}
\end{figure}

It is worth noting that for the first query, we cannot build the click user profile because there is no previously clicked document. However, we can still build the query user profile for the first query. We show the improvement in performance of the personalised methods over \emph{Adeyanju's} in term of MAP metric with different query positions in Figure \ref{img1}. Here the statistical significance is verified by t-test (p < 0.01). For the first query within a search session, our method, which can use only the query user profile, significantly improves the query suggestion performance over \emph{Adeyanju's}. It again confirms the effectiveness of the query information on personalised query suggestion for Intranet search.

From the second query, we can build both the query and click user profiles. One can see that the higher position of a query is, the larger improvement in performance the personalised query suggestion can be achieved. Specifically, from the query with high positions (i.e., $\geq 4$), the improvements of \emph{Click} and \emph{Ours} are 11.37\% and 18.22\%, respectively.  Figure \ref{img1} also shows that \emph{Ours} outperforms \emph{Click} significantly with the improvements of at least 3.45\% (p < 0.01). It indicates that richer user profiles (by observing more clicked documents and submitted queries during the search session) help achieve better query suggestion performances. The findings offer future research directions that use user profiles which go beyond single sessions

\subsection{Performance on Different Query Lengths}
The query length is defined by the number of words in the query (e.g., the query ``University webmail'' has the length of 2). The length of a query might give an indication as to how specific the information need of an individual user is (i.e., a longer query can typically be assumed to reflect a more specific information need). In this experiment, we aim to show the impact of personalisation on query suggestion with different query lengths. We label each query by its length, which is the number of words in the query (i.e., from one to three and $\geq 4$ words because there are few queries containing more than three words (i.e., 5.7\% of queries in the query logs)).

\begin{figure}
\centering
\includegraphics[width=6.0cm]{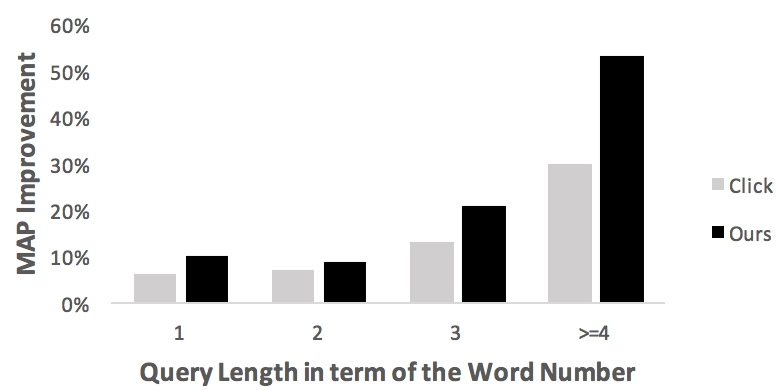}
\caption{Relative MAP improvements over \emph{Adeyanju's}  with different query lengths.}
\label{img2}
\end{figure}

Figure \ref{img2} shows the improvement in performance of \emph{Click} and \emph{Ours} over \emph{Adeyanju's} in term of MAP with different query lengths. We see that personalisation methods achieve significantly better performances than the non-personalised method does (p < 0.01). Even for short queries (length 1 and 2) which tend to be more generic, the \emph{Click} and \emph{Ours} methods outperform \emph{Adeyanju's} method with the improvement of more than 6.11\% and 9.09\%, respectively. We see that the longer a query is, the higher improvement personalised methods can achieve. Specifically, with a longer query (i.e., with length $\geq 4$), the \emph{Click} and \emph{Ours} methods yield the highest improvements, i.e., 29.82\% and 53.85\%, respectively. This indicates that a longer query would also get more benefit from personalisation.

Figure \ref{img2} also indicates that by combining the query user profile with the click user profile, \emph{Ours} significantly improves the query suggestion performance over \emph{Click} (p < 0.01). Moreover, the improvements are larger with the longer queries (i.e., with length $\geq 3$). In particular, the improvements of \emph{Ours} over \emph{Click} are 6.7\% and 18.2\% on the query length 3 and the query length $\geq 4$, respectively.
\section{CONCLUSIONS}
\label{sec5}
In this paper, we proposed a personalised query suggestion framework and showed how it performed on Intranet search. We built two session-specific temporal user profiles, a query user profile using the submitted queries, and a click user profile using the clicked documents. We then extracted the personalised features using the two profiles and combined them with non-personalised features to learn a ranking model using LamdaMART. Finally, we used the ranking model to re-rank the query suggestion list returned by a state-of-the-art query suggestion approach for Intranet search.  

Experimental results on a large-scale query log dataset collected from a university intranet search engine show that personalisation significantly improved the query suggestion performance. Using both the click user profile and query user profile achieved the highest performance indicating that personalised query suggestion for Intranet search should take into account both click and query information. Moreover, the positive impact of personalised query suggestions is more pronounced with longer queries and queries submitted later within a session.


\bibliographystyle{abbrv}
{\small \bibliography{sigproc}}
\end{document}